\documentclass{article}
\usepackage{psfig}
\input epsf 
\usepackage{emulateapj}
\usepackage{apjfonts}

\newcommand{\half}{\frac{1}{2}}
\newcommand{\Te}{T_\mathrm{e}}
\newcommand{\Td}{T_\mathrm{d}}
\newcommand{\Tm}{T_\mathrm{m}}
\newcommand{\ri}{r_\mathrm{i}}
\newcommand{\rc}{r_\mathrm{c}}
\newcommand{\rms}{r_\mathrm{ms}}
\newcommand{\me}{m_\mathrm{e}}
\newcommand{\taue}{\tau_\mathrm{e}}
\newcommand{\source}{GRS~1915+105 }

\slugcomment{Submitted to The Astrophysical Journal, Letters}

\begin{document}

\title{The Compton Microscope: Using the Energy Dependence of QPO Amplitudes \\to Probe Their Origin in Accretion Disks}

\author{Dana E. Lehr\altaffilmark{1}, Robert V. Wagoner\altaffilmark{1}, 
J\"{o}rn Wilms\altaffilmark{2} }

\altaffiltext{1}{Dept.\ of Physics, Stanford University, Stanford, 
CA 94305--4060; \{danalehr, wagoner\}@stanford.edu}

\altaffiltext{2}{Institut f\"{u}r Astronomie und Astrophysik,
Abt.\ Astronomie, Waldh\"{a}user Str.\ 64, D--72076
T\"{u}bingen, Germany; wilms@astro.uni-tuebingen.de }

\begin{abstract}

We report the development of a new tool to determine the origin of
quasi-periodic oscillations (QPOs) in accretion disk systems. The technique
uses the source energy spectrum and the energy dependence of the QPO
fractional amplitude to restrict the location of the emission region of the
modulated photons, which are assumed to originate in the inner accretion
disk. Both Monte Carlo and semi-analytical methods are presented. We assume
the accretion disk is enshrouded by a slab atmosphere of hot electrons in
which unsaturated Compton scattering produces the high-energy
spectrum. Properties of the atmosphere, in particular the electron
temperature, are assumed functions of radius from the central compact object.
We show that our model reproduces the observed energy dependence of the
fractional amplitude of the 67\,Hz QPO in \source if the QPO is assumed to
originate within five gravitational radii from the innermost stable circular
orbit.

\end{abstract}


\section{Introduction}\label{sect:intro}

Quasi-periodic oscillations (QPOs) have been seen in the power spectra of
X-ray binaries since work with the EXOSAT and \textsl{Ginga} satellites (van der Klis 
1995)\markcite{V}. With the launch of the Rossi X-Ray Timing
Explorer (RXTE) satellite in 1995, advances in instrumentation have allowed observers
to measure QPOs at such high frequencies that their origin has been associated
with the inner regions of the accretion disks which are expected to exist in
binary systems. Evidence on the exact location of the QPO generation, however,
remains unclear. Observed properties of the QPOs, such as centroid frequency,
width, amplitude, phase lag, and coherence, have been investigated for insight
into the physical processes which produce these oscillations.  Another
characteristic of QPOs, the energy dependence of their fractional
root-mean-square (rms) amplitude, remains as a relatively unexplored tool to
diagnose their origin.  In this Letter we report the development of a tool
which utilizes the observed energy dependence of the QPO amplitude to locate
the accretion disk region in which the modulation may occur.

Analyses of QPO mechanisms must be commensurate with a model for the overall
energy spectrum of the source. In particular, a frequent component of X-ray
binary spectra is a high-energy power-law, usually interpreted to arise from
unsaturated inverse Comptonization by hot electrons. These electrons are
typically associated with a high-temperature atmosphere or corona for which
the geometry, production, and physical properties are uncertain (Dove, Wilms,
\& Begelman 1997, and references therein)\markcite{DWB}. Here we demonstrate a
method for localizing the QPO emission site using the observed energy
dependence of the QPO amplitude and the observed source energy spectrum. We
assume a particular form of a Comptonizing atmosphere which adequately fits
the observed source spectrum. Ideally one would like to constrain the
atmosphere's true structure by employing additional observed properties of the
QPO, but this is beyond the scope of this Letter.

\section{Comptonizing Model}\label{sect:model}

To demonstrate the qualitative effects of a Comptonizing atmosphere on the
energy dependence of QPO fractional amplitudes we consider a simple model of
an accretion disk enshrouded by a steady-state, locally plane-parallel
corona. The corona consists of thermalized, nonrelativistic hot electrons
which are optically thin to absorption and have no bulk motion. In addition,
properties of the atmosphere, such as the electron temperature $\Te$, may vary
slowly in the radial direction along the surface of the accretion
disk. Photons originate in the cold accretion disk, which has temperature
profile $\Td(r)$, and scatter through the slab atmosphere before escaping the
cloud or re-entering the disk to be absorbed. Here we do not consider the
effects of strong gravity on photon paths.

A measure of the distortion of a photon spectrum by Compton scattering through
a nonrelativistic thermal distribution of electrons is given by the Compton
$y$ parameter,
\begin{equation} 
y = \frac{4k\Te}{\me c^2} \max(\taue,\taue^2), 
\label{eq:ypar}
\end{equation}
where $\me c^2$ is the electron rest energy and $\taue$ is the optical depth
of the cloud. If we restrict our cloud parameters such that $y(r) \ll 1$, then
the spectral profile of the observed photon energy $E$ is described in the
range $k\Td \ll E \ll k\Te$ by unsaturated Comptonization of a soft photon
input (Rybicki \& Lightman 1979)\markcite{rl}. The emergent photon flux
density in this regime is given by a power-law dependence,
\begin{equation}
dN/dE \propto E^{-\alpha},
\end{equation}
where 
\begin{equation}
\alpha =-\half + \left(\frac{9}{4}+\frac{4}{y}\right)^\half \approx
2y^{-\half} - \half + {\mathcal O}(y^\half), 
\label{eq:powlaw}
\end{equation}
and $N$ has units of photons\,cm$^{-2}$\,s$^{-1}$.  

If we choose the radial gradient of the electron temperature such that $y$
decreases with $r$, then it follows from
equations~\mbox{(\ref{eq:ypar})--(\ref{eq:powlaw})} that
$\alpha(r_1)<\alpha(r_2)$ if $r_1<r_2$. That is, simply, photons which
originate near the inner edge of the accretion disk are more strongly
Comptonized than photons which originate in the outer regions of the disk. If
we apply the above to a QPO which originates generally inside the radius at
which the maximum (unmodulated) luminosity per unit radius is emitted from the
disk, the QPO fractional amplitude can be an increasing function of energy.
More particularly, the QPO amplitude may continue to increase over a domain
which includes energies that are too high to be traditionally associated with
``thermal'' processes from the relatively cold disk.

\section{Monte-Carlo Simulations}\label{sect:mc}

One approach to quantitatively assess the above approach is to conduct Monte
Carlo simulations which numerically propagate accretion disk photons through a
Comptonizing corona. In our simulations we adopt the standard
semi-relativistic thin disk model of Shakura \& Sunyaev (1973)\markcite{SS}
with the modified Newtonian potential (Nowak \& Wagoner 1991)\markcite{NW}
\begin{equation}
\Phi = -(GM/r)[1-3(GM/rc^2)+12(GM/rc^2)^2]\, ,
\end{equation}
where $M$ is the mass of the central black hole.
With this choice of potential the disk temperature is given by (Nowak 1992)\markcite{N}
\begin{eqnarray}
\hspace{-.3cm}\Td(r) = 10.286\,\Tm\,r^{-\frac{3}{4}}
\!\left(1-\frac{8}{r}+\frac{60}{r^2}\right)^{\!\!\frac{1}{4}}
\!\!\left(1-\sqrt{\frac{6}{r-6+36/r}}\right)^{\!\!\frac{1}{4}}\!\!,
\end{eqnarray}
where the radial coordinate is now in units of $GM/c^2$ and the maximum
temperature $\Tm$ of the disk is reached at the radius $r=9.795$. The
electron temperature of the atmosphere is taken to be 
\begin{equation}
\Te(r) = T_0(r/\ri)^{-1} ,
\end{equation}
where $\ri$ is the inner radius of the disk and $T_0$ is a constant. We assume
the accretion disk extends down to the innermost stable circular orbit $\rms$
such that $\ri = \rms = 6$ for our assumed slowly rotating black hole. To
completely specify a model spectrum (minus an overall normalization), we must
choose $\Tm,\ T_0,$ and the optical depth $\tau$ of the electron
cloud. Finally, an additional parameter $\rc$ depicts the outer radius of the
electron cloud, which is not constrained to cover the entire disk.

The simulation code for the slab model is based on the ``particle escape
weighting'' algorithm presented by Pozdnyakov, Sobol, \& Sunyaev
(1983)\markcite{PSS}. A brief description of the central driver of the code
has been given by Nowak et al.\ (1999)\markcite{NWVB}. Photons are created with
a disk-blackbody energy distribution and undergo repeated Compton scattering
until escape from the atmosphere or absorption into the disk. The initial
angular distribution of the photons is chosen for a scattering-dominated
atmosphere (Chandrasekhar 1960)\markcite{SW}\markcite{CH}. We employ the
relativistic scattering formulae as well as the differential Klein-Nishina
cross section to compute the scattering angle at each collision. In addition,
the electrons are taken to have a relativistic Maxwellian distribution at
the local temperature of the atmosphere. 

To simplify the simulations the slab atmosphere is divided into approximately
600 bins (corresponding to radial annuli in the accretion disk), each of which
is given constant disk and electron temperatures equal to the respective local
temperatures at the center of the annulus. An equal number of photons enter
each bin at the base of the atmosphere and scatter within a uniform
temperature cloud. The escaped photons in each bin are appropriately weighted
by a factor $\Td^3 r \Delta r$ and summed to create the final energy spectrum
of the model. A similar integration over a finite radial range calculates the
spectrum of the localized disk region which is hypothesized to produce the QPO
photons. The ratio of the spectrum of the modulated photons to the total disk
spectrum is then proportional to the fractional rms amplitude of the QPO as a
function of energy.

We note that we do not include reflection off of the cold disk in these
exploratory calculations. Rather, we assume that the absorbed photons will be
thermalized by the disk and re-emitted with a blackbody distribution at the
local disk temperature. That is, we treat the disk as a perfect blackbody.
Photon number is not conserved in this process, but the total energy flux
deposited into the disk is assumed to be re-emitted locally. Since this
re-emitted distribution is the same as our initial input distribution, we can
renormalize the escaped photon energy flux to include the total energy flux
lost to absorption in the disk. We apply this rescaling of the escaped flux at
each annular radius before we add the photons to compute the total energy
spectrum for the model.

\section{Application to \source}\label{sect:appl}

Three known black hole candidates (BHCs) are each observed to display a
high-frequency QPO (Morgan, Remillard, \& Greiner 1997; Remillard et al.
1999a,b)\markcite{MRG}\markcite{R1550}\markcite{R1655}; two of these, a 67\,Hz
feature in \source and a 300\,Hz feature in GRO J1655-40, seem to remain
stable in frequency despite appreciable changes in the source luminosity. The
origin of these QPOs has been suggested to invoke effects of general
relativity in the inner accretion disk since several time scales there can be
associated with fast QPO at X-ray energies. Proposals for the QPO origins
include: hot blobs at the innermost stable orbit (Morgan et al.\ 1997),
frame-dragging effects (Merloni et al.\ 1999; Cui, Zhang, \& Chen 1998;
Markovic \& Lamb 1998)\markcite{MVSB}\markcite{CZC}\markcite{ML},
inertial-acoustic instabilities (Milson \& Taam 1997; Honma, Matsumoto, \&
Kato 1992)\markcite{MT}\markcite{HMK}, and diskoseismic oscillations (Nowak et
al.\ 1997; Perez et al.\ 1997)\markcite{NWBL}\markcite{PSWL}. To our knowledge
the consistency of these models with the energy dependence of the QPO
fractional amplitude has never been illustrated quantitatively.

To demonstrate the application of our tool we fit our numerically simulated
energy spectra to data from RXTE observations of \source taken during early
1996 when the 67\,Hz QPO was particularly strong (1996 May 05; Morgan et al.\
1997)\markcite{MRG}. In our analysis we limited the energy range of the
Standard-2 PCA data from 2.5 to 25\,keV and imposed a 1\% systematic error
over the entire energy range (Wilms et al.\ 1999)\markcite{WND}. Spectral
fitting was done using a grid of our four-parameter Monte Carlo spectra as an
additive table model in XSPEC, version 10.00 (Arnaud 1996)\markcite{A}. The
attenuation of radiation by the interstellar medium was taken into account
using a second multiplicative model which employs the cross-sections of Ba\l
uci\'{n}ska-Church \& McCammon (1992)\markcite{BCM}.

The Monte Carlo model spectrum which best fits the data is shown in
Figure~\ref{fig:xspec} with residuals plotted as the ratio of the data 
to the model. For comparison we also present the residuals of the
best-fit spectrum from the phenomenological model of an absorbed
disk-blackbody plus an additional blackbody component. Fit parameters for both
models are listed in Table~\ref{table:xspec}.

We note that the residuals for both models indicate a probable iron line near
6.4\,keV.  An \textsl{ad hoc} addition of such a component \\

\smallskip
\centerline{
\epsfxsize=0.90\hsize {\epsfbox{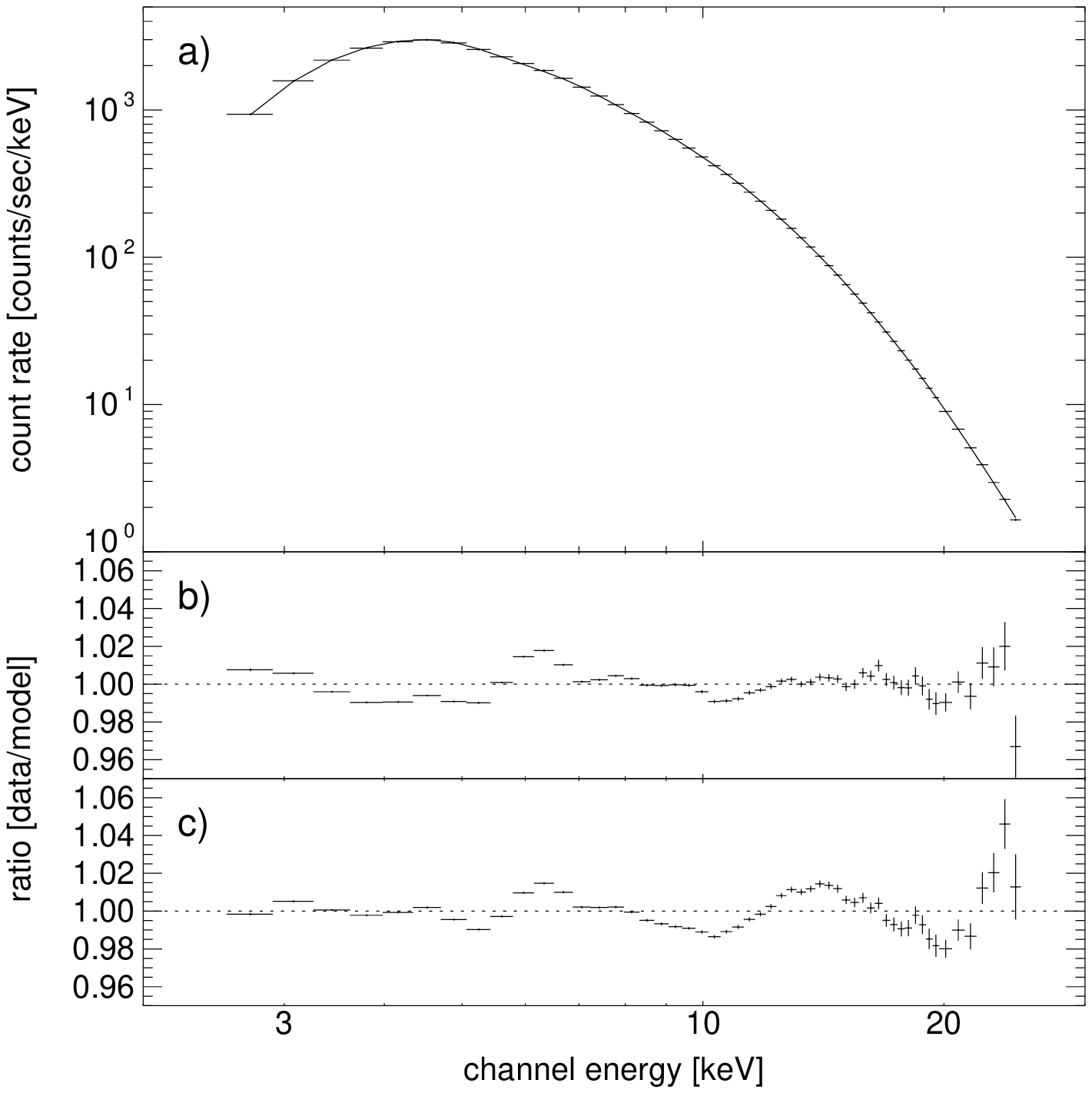}}
}
\figcaption{Spectral modeling of PCA data from RXTE observations
of \source on 1996 May 05. (a) Count rate spectrum and the best-fit Monte Carlo
model spectrum. (b) Residuals from the Monte Carlo model shown as the ratio
of the data to the model. (c) Residuals from the best-fit disk-blackbody plus
blackbody model. \label{fig:xspec}}
\bigskip

\begin{deluxetable}{lccccccccc}
\tablecaption{Parameters for best-fit XSPEC models}
\tablehead{
\colhead{Model}&
\colhead{nH}&
\colhead{$k\Tm$}&
\colhead{$kT_0$}&
\colhead{$\tau$}&
\colhead{$\rc$}&
\colhead{$kT_\mathrm{in}$}&
\colhead{$kT$}&
\colhead{$\chi^2/\mathrm{dof}$}&
\colhead{$\chi^2_{\nu}$} \\
&($\times 10^{22}$)&(keV)&(keV)&&($GM/c^2$)&(keV)&(keV)&&
}
\startdata
Monte Carlo&$5.07^{+ 0.08}_{- 0.05}$&$1.61^{+ 0.01}_{- 0.01}$&$5.12^{+ 0.08}_{- 0.07}$&$6.58^{+ 0.27}_{- 0.26}$&$20.2^{+ 0.98}_{- 0.42}$&\nodata&\nodata&$24.5/48$&$0.51$\\[.1cm]
diskbb + bbody&$4.57^{+ 0.14}_{- 0.14}$&\nodata&\nodata&\nodata&\nodata&$1.65^{+ 0.02}_{- 0.02}$&$2.64^{+ 0.03}_{- 0.02}$&$46.8/49$&$0.96$\\
\enddata
\tablecomments{Best-fit model parameters used in spectral modeling of
PCA data from RXTE observations of \source on 1996 May 05. Parameters for the
Monte Carlo model are described in text. Parameters for the disk-blackbody plus
blackbody model are: temperature at innermost disk radius $T_\mathrm{in}$ and
blackbody temperature $T$. \label{table:xspec}}
\end{deluxetable}

\noindent slightly improves
the residuals from 5--10\,keV ($\Delta \chi^2 = 0.39$) but does not
significantly change the fit parameters for either model. Since the line is
unimportant for both the overall flux and the resulting QPO energy dependence,
we did not include it in our further modeling. 

Using the best-fit parameters for the Monte Carlo model, we calculated the
ratios of the modulated photon spectra to the total disk spectrum for three
hypothesized regions of the 67\,Hz QPO production. The resulting energy
dependences of the QPO fractional amplitude are compared to the measurement of
Morgan et al.\ (1997)\markcite{MRG} in Figure~\ref{fig:QPOamplitude}.  Each of
the curves in Figure~\ref{fig:QPOamplitude} has been normalized by a
multiplicative factor to best describe the observational data. (This
normalization factor corresponds to the fraction of photons emitted from the
disk region which are modulated by the QPO mechanism.) Two of the curves are
an acceptable fit to the data: one depicting QPO photons emitted from
6--7\,$GM/c^2$ and one emitted from 7.5--8.5\,$GM/c^2$. The third example,
depicting a QPO produced in an outer disk region from 11--12\,$GM/c^2$, does
not reproduce the observation.

\section{Analytic Approximation}\label{sect:analyt}

Wagoner \& Silbergleit (1999)\markcite{WS} have developed an analytical
formalism which considers Compton scattering in the slab model above for an
assumed angular distribution of the photons (which is valid for the diffusion
approximation and the two-stream approximation). Using the local approximation
that gradients normal to the disk surface are dominant, they solve the
Boltzmann equation governing the evolution of the radiation field under
collisions (i.e., scattering events in the atmosphere). They invoke the
Kompaneets approximation that the change in the photon energy in each
scattering event is much less than $k\Te$ and neglect all other photon
processes (such as bremsstrahlung). Finally, they neglect stimulated
scattering; that is, the photon occupation number is assumed to be small.

Under these approximations the authors obtain an analytical expression for
the photon energy flux (per unit energy, perpendicular to the disk surface) at
any optical depth in the atmosphere. The solution involves only integrals over
the input flux, which is a specified function of energy and may be a slowly
varying function of $r$. For an input flux which is radially dependent, the
emergent flux may be summed over radii if the physical thickness of the
atmosphere is much less than the scale of radial gradients of the disk or
cloud properties (a restriction which we note applies to our Monte Carlo
model, as well).

Wagoner \& Silbergleit apply their formalism to thin accretion disk
atmospheres. Employing this application, we can compute the energy spectrum of
QPO photons relative to the total energy spectrum of the disk and compare with
the Monte Carlo application above. The parameters of the analytical model are
the same as in our numerical work, namely $\Tm,\ T_0,\ \tau,$ and $\rc$. (We
note that the formalism allows a more general application in which the inner
radius of the disk exceeds $\rms$ and in which the radial dependence of the
electron temperature is arbitrary. The optical depth $\tau$ of the atmosphere
may also be a slowly-varying function of $r$.) To complete the model one must
specify the boundary condition at the top of the atmosphere ($\tau = 0$); the
outer boundary condition was taken from standard scattering atmospheres
(Chandrasekhar 1960).\markcite{CH}

The analytic approximation was employed within a proof-of-principle
calculation to investigate the sensitivity of this tool. Over a range of
parameters involving lower optical depths and higher cloud temperatures than
the above fit to the 1996 May 05 observation of \source, we compared our Monte
Carlo results to the analytical predictions. The common results were in good
agreement. In particular, we obtained similar sensitivity to the location of
the injected modulated photons.

\section{Summary}\label{sect:disc}

We have demonstrated a tool to locate the origin of QPO photons in accretion
disk systems; the method uses the observed energy spectrum and the observed
energy dependence\\

\bigskip
\centerline{ 
\epsfxsize=0.90\hsize {\epsfbox{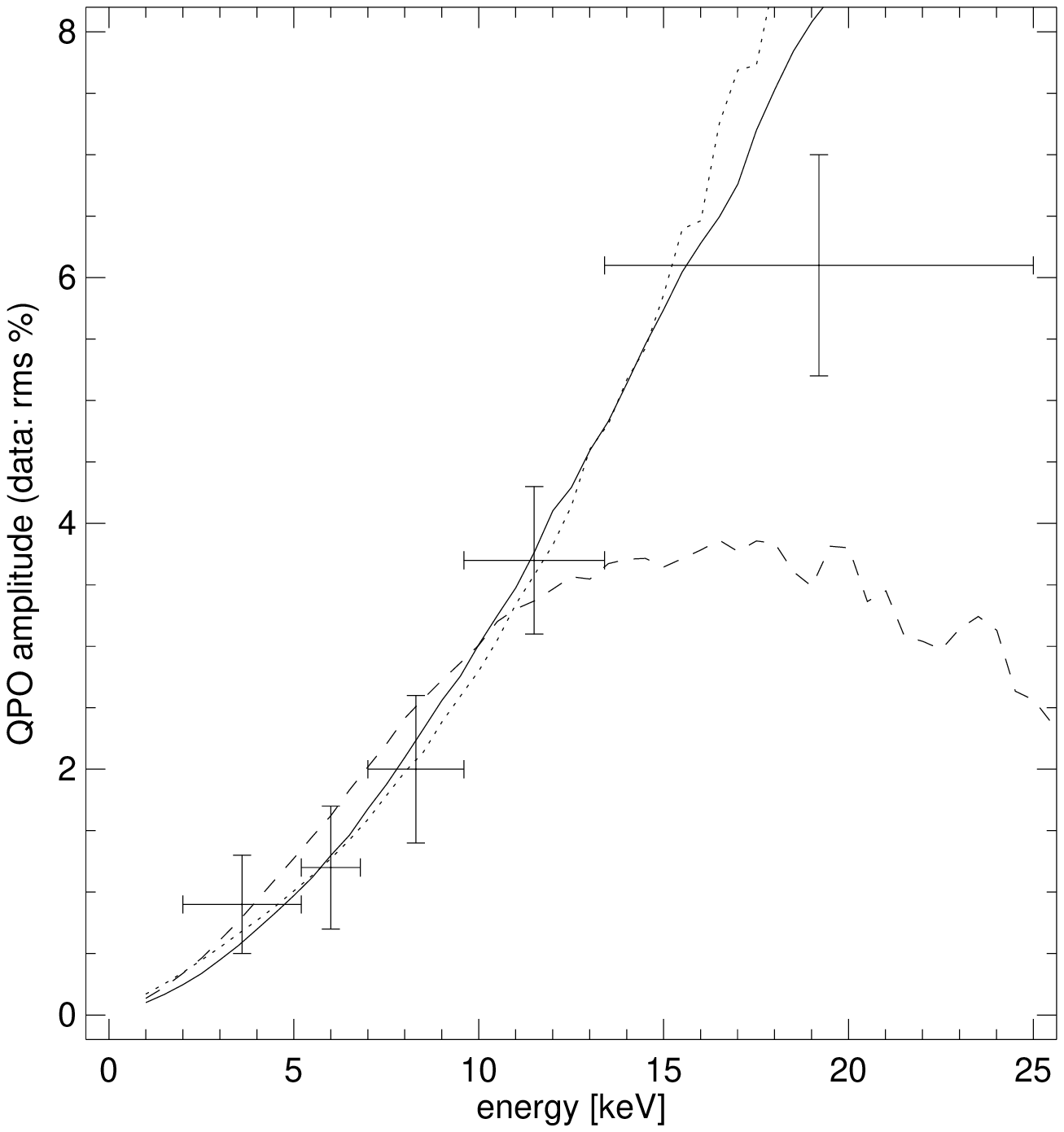}} 
}
\figcaption{Energy dependence of the 67\,Hz QPO of \source on 1996 May 05.
Monte Carlo simulations are compared to data (5 points) from Morgan et al.\
(1997). \textit{Dotted:} ratio of spectrum from photons injected at 6--7
$GM/c^2$ to total disk spectrum; ratio is normalized by a factor of
$n=2.78$. \textit{Solid:} same ratio for photons injected at 7.5--8.5
$GM/c^2$ ($n=0.50$). \textit{Dashed:} same ratio for photons
injected at 11--12 $GM/c^2$ ($n=0.34$). 
\label{fig:QPOamplitude}}
\bigskip

of the QPO fractional amplitude to constrain the disk location in which the
photon modulation may occur. Applying this technique to the BHC \source, we
show that the functional energy dependence of the QPO amplitude constrains the
origin of the QPO to be between the inner disk edge and just outside the
maximum in the radial epicyclic frequency at 8\,$GM/c^2$. This location is
consistent with the models described in \S\ref{sect:appl}. From experience
with our Monte Carlo and analytic models we can also suggest that observations
of \source which require hotter clouds to model comparatively harder spectra
(e.g.\ 1996 June 12 from Morgan et al.\ 1997) may more precisely constrain the
location of the QPO production.


The natural development of this tool will incorporate other geometries, as
well as different temperature and density structures, to model the
Comptonizing atmosphere. In addition, one must relax the assumption of a
slowly-rotating central black hole. Although here we model the energy spectrum
of \source adequately with our simple choice of coronal construction, we do
not fully constrain the structure of the atmosphere using relevant
observations such as timing or coherence. The latter data should be used in
conjunction with energy spectra to reveal the true nature of the electron
clouds which might produce the high-energy photons seen in many BHC. The tool
discussed in this Letter could then be a powerful indicator of the origin of
QPOs in accretion disk systems.

\acknowledgements

The authors would like to acknowledge innumerable useful conversations with
Mallory Roberts.  We also thank Keith Arnaud for his assistance with the XSPEC
table model file format. This work was supported by NASA Graduate Student
Researchers Program grant NGT 5-50044 to D.E.L., NASA grant NAG 5-3102 to
R.V.W., and grant number Sta 173/22 of the Deutsche Forschungsgemeinschaft to
J.W. R.V.W. and J.W. also thank the Aspen Center for Physics for support
during a 1999 summer workshop. This research has made use of data obtained
through the High Energy Astrophysics Science Archive Research Center Online
Service, provided by the NASA/Goddard Space Flight Center.





%

\end{document}